%% file: main.tex
\documentclass[sigconf,screen]{acmart}
\AtBeginDocument{%
  }

\setcopyright{acmlicensed}
\copyrightyear{2025}
\acmYear{2025}
\acmDOI{XXXXXXX.XXXXXXX}
\acmBooktitle{Companion Proceedings}
\acmISBN{978-1-4503-XXXX-X/18/06}

\usepackage{pifont}
\usepackage[normalem]{ulem}
\usepackage{fontawesome}
\usepackage{threeparttable}
\usepackage{enumitem}

\newcommand{\ours}[1]{\textsc{KGMAF}}

\def\eg{\textit{e.g.,} }
\def\ie{\textit{i.e.,} }

\begin{document}

\title{Knowledge-Guided Multi-Agent Framework for Automated Requirements Development: A Vision}


\author{Jiangping Huang}
\email{jiangpyeong@gmail.com}
\affiliation{%
  \institution{Chongqing University of Posts and Telecommunications}
  \state{Chongqing}
  \country{China}
}
\email{jiangpyeong@gmail.com}

\author{Dongming Jin}
\affiliation{%
  \institution{Peking University}
  \city{Beijing}
  \country{China}}
\email{dmjin@stu.pku.edu.cn}

\author{Weisong Sun}
\affiliation{%
 \institution{Nanyang Technological University}
 \country{Singapore}}
\email{weisong.sun@ntu.edu.sg}

\author{Yang Liu}
\affiliation{%
 \institution{Nanyang Technological University}
 \country{Singapore}}
\email{yangliu@ntu.edu.sg}

\author{Zhi Jin}
\affiliation{%
  \institution{Peking University}
  \city{Beijing}
  \country{China}}
\email{zhijin@pku.edu.cn}

\renewcommand{\shortauthors}{J. Huang, D. Jin, W. Sun, Y. Liu, Z. Jin.}

\input{sections/abstract}

\ccsdesc[500]{Software and its engineering~Requirements analysis}
\ccsdesc[500]{Computing methodologies~Natural language generation}
\ccsdesc[500]{Computing methodologies~Multi-agent planning}
\ccsdesc[500]{Computing methodologies~Multi-agent systems}

\keywords{Automated Requirements Development, Requirements Engineering, Large Language Models, Multi-Agent Systems}


\maketitle

\input{sections/introduction}
\input{sections/framework}
\input{sections/casestudy}

\input{sections/future_work}

\section{Conclusion}
\label{sec:06_conclusion}
This paper envisions a knowledge-guided multi-agent framework named \ours{} for automated requirements development. \ours{} 
aims to address gaps in current automation systems for SE, which prioritize code development and overlook the complexities of requirements tasks. \ours{} is composed of six specialized agents and an artifact pool to improve efficiency and accuracy. Specifically, \ours{} outlines the functionality, actions, and knowledge of each agent and provides the conceptual design of the artifact pool. Our case study highlights the potential of KGMAF in real-world scenarios. Finally, we outline several research opportunities for implementing and enhancing automated requirements development using multi-agent systems. We believe that {\sc KGMAF} will play a pivotal role in shaping the future of automated requirements development in the era of LLMs.

\bibliographystyle{unsrt}
\bibliography{ref}

\end{document}

%% file: sections/abstract.tex
\begin{abstract}
Requirements Engineering (RE) is an initial and critical phase in the software development process. It aims to produce a well-defined software requirements specification from rough ideas. This process involves a series of tasks (\eg elicitation and analysis) and roles (\eg the interviewer and analyst), which is a time-consuming and human-intensive process. Recently, large language models (LLMs) and multi-agent systems have been proven successful in the software development process (\eg MetaGPT and ChatDev). However, they primarily focus on code development and lack the design of agents specifically for RE. To address this gap, we envision a knowledge-guided multi-agent framework named \ours{} for automating the requirements development process. \ours{} is composed of six agents and an artifacts pool. Each agent is equipped with a functionality, predefined actions, and knowledge to perform various requirements tasks. The artifacts pool is designed to store the intermediate artifacts generated by these agents to facilitate their collaboration. We conducted a case study to validate the practicality of our \ours{}. This vision paper highlights the design of the multi-agent system for RE, with preliminary results from our ongoing work and an agenda for
future research.
\end{abstract}

%% file: sections/introduction.tex
\section{Introduction}
\label{sec:01_introduction}

\begin{figure*}[t]
    \centering
    \includegraphics[width=\linewidth]{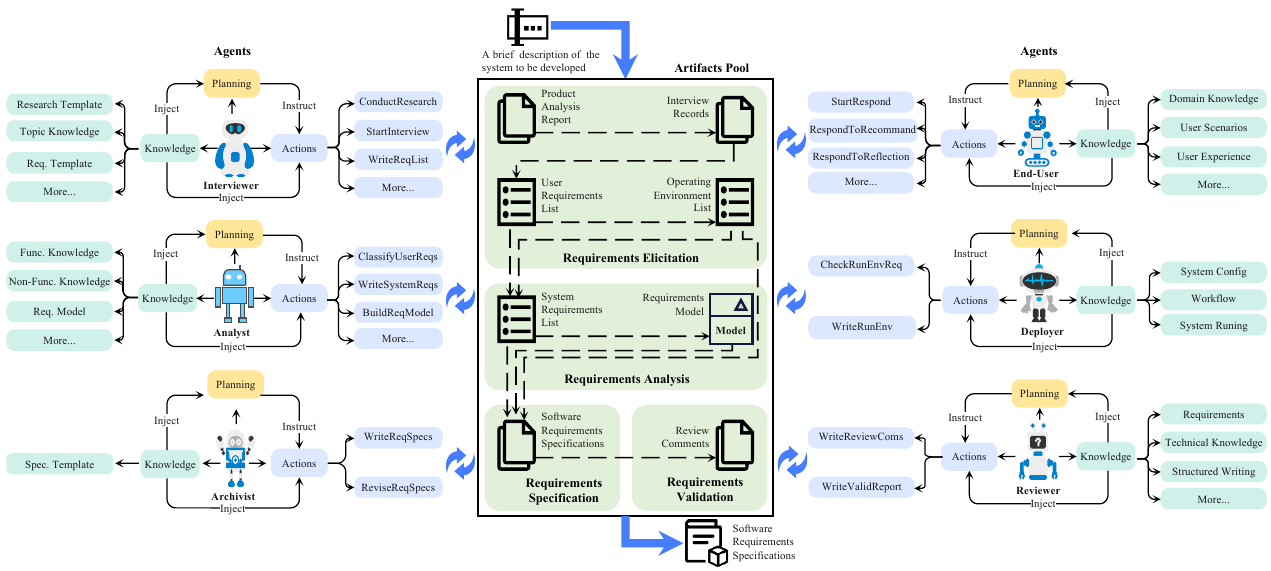}
    \vspace{-6mm}
    \caption{An overview of the knowledge-guided multi-agent framework for automated requirement development.}
    \label{fig:figure_01}
\end{figure*}

The success of a software system depends on whether it can meet the needs of its users and its environment~\cite{cheng2007research, nuseibeh2000requirements, parnas1999software}. 
Requirements Engineering (RE), a pivotal phase in the software development lifecycle~\cite{jin2024mare}, involves deriving and determining appropriate, achievable user and environmental requirements~\cite{jinbooks}. 
RE involves a series of tasks, including obtaining user requirements, understanding the operating environment, analyzing requirements, archiving requirements specifications, and validating requirements quality. These tasks need to involve multiple roles and will cost lots of time and human effort.

Recently, large language models (LLMs) have shown remarkable capabilities in various individual software engineering tasks, \eg requirements modeling~\cite{jin2024evaluation} and code summarization~\cite{2025-LLM4CodeSum}. Simultaneously, autonomous agents~\cite{zhou2023webarena} built on LLMs offer opportunities to replicate human workflow and perform end-to-end software development. For example, ChatDev~\cite{qian2024chatdev} assigns different roles to agents, with each role acting as a coordinator to automatically fulfill user requirements through a chat chain. However, it focuses on visual design, code generation, and execution and does not involve requirements development. MetaGPT~\cite{hong2023metagpt} integrates efficient human workflows into multi-agent collaboration, covering the entire development process. However, its requirements analysis is limited to generating requirements documents and lacks support for requirements elicitation and other related activities.

Thus, existing works~\cite{qian2024chatdev, hong2023metagpt, web:gpt-engineer} on multi-agent systems for software development primarily focus on code development and lack a sufficient design for the requirements development. Additionally, the roles in software development require a variety of knowledge to support development activities. Current works~\cite{qian2024chatdev, hong2023metagpt, web:autogpt} just directly assign agents to act as roles and do not clearly define the software engineering knowledge they need.

To address this gap, this paper envisions a knowledge-guided multi-agent framework named \ours{} for automating the requirements development process. Specifically, given a brief idea of a software, \ours{} autonomously performs the tasks of RE, \ie elicitation~\cite{GorerA24}, analysis~\cite{AbdeenUWCSG24}, specification~\cite{KrishnaGVJ24}, and validation~\cite{veizaga2024automated}, to achieve end-to-end software requirements specifications (SRS) and other intermediate artifacts generation. \ours{} is composed of six agents, \ie interviewer, end-user, analyst, archivist, reviewer, and deployer. Each agent is equipped with the functionality, predefined actions, and knowledge to plan and perform tasks. Besides, inspired by blackboard mechansim~\cite{craig1988blackboard}, \ours{} has an artifacts pool to store the requirements artifacts generated by these agents. Meanwhile, the state of the artifacts pool determines the workflow and collaboration among agents. Each agent continuously monitors the pool's state (\eg addition and update) and plan-then-execute actions based on the state. To validate the practicality of \ours{}, we conduct a case study on a real-world insurance management system~\cite{webcase} and present our preliminary results (Figure \ref{fig:case study}). Finally, we discuss an agenda for future research on multi-agent framework for RE. We hope that our vision paper can facilitate intelligent RE in the era of LLMs and multi-agent systems.

%% file: sections/framework.tex
\section{Conceptual Design of \ours{}}
\label{sec:03_kdre}

\subsection{Overview}

The goal of automated requirements development is to produce a well-defined SRS from a brief description of the system to be developed. Automated requirements development encompasses four critical activities (\ie elicitation, analysis, specification, and validation). In \ours{}, these activities are completed collaboratively by six LLM-based agents (\ie the Interviewer, End-User, Analyst, Deployer, Archivist, and Reviewer) and an artifacts pool. Each agent is equipped with specific knowledge, including both professional expertise and domain knowledge, to formulate a plan and perform a series of actions. They work in a pipeline as shown in Figure \ref{fig:figure_01}. We describe the details, including the setting and design of our \ours{}\footnote{The "More..." indicates that additional specific knowledge and actions are available.}, in the following sections.

\subsection{\ours{} Setting}
\textbf{Agents Setting.} Each agent is responsible for specific requirements tasks as its functionality and consists of three modules, \ie predefined actions, planning mechanism, and required knowledge. The predefined actions are the capabilities assigned to each agent. The planning mechanism is to think and select the next action from the predefined actions. The performance of both modules requires an injection of requirements-related knowledge, which can be achieved by designing complex prompts with various strategies~\cite{wei2022chain, yao2024tree}.

\noindent \textbf{The Artifacts Pool Setting.} The artifacts pool is designed to store requirements artifacts produced by agents. The status of this pool plays a crucial role in driving collaboration among the agents. Specifically, each agent continuously monitors the state of the artifact pool. When an artifact (\eg requirements model) is added or updated to the pool, some agents will be triggered to plan and perform corresponding actions.

\begin{figure*}[t]
    \centering
    \includegraphics[width=\linewidth]{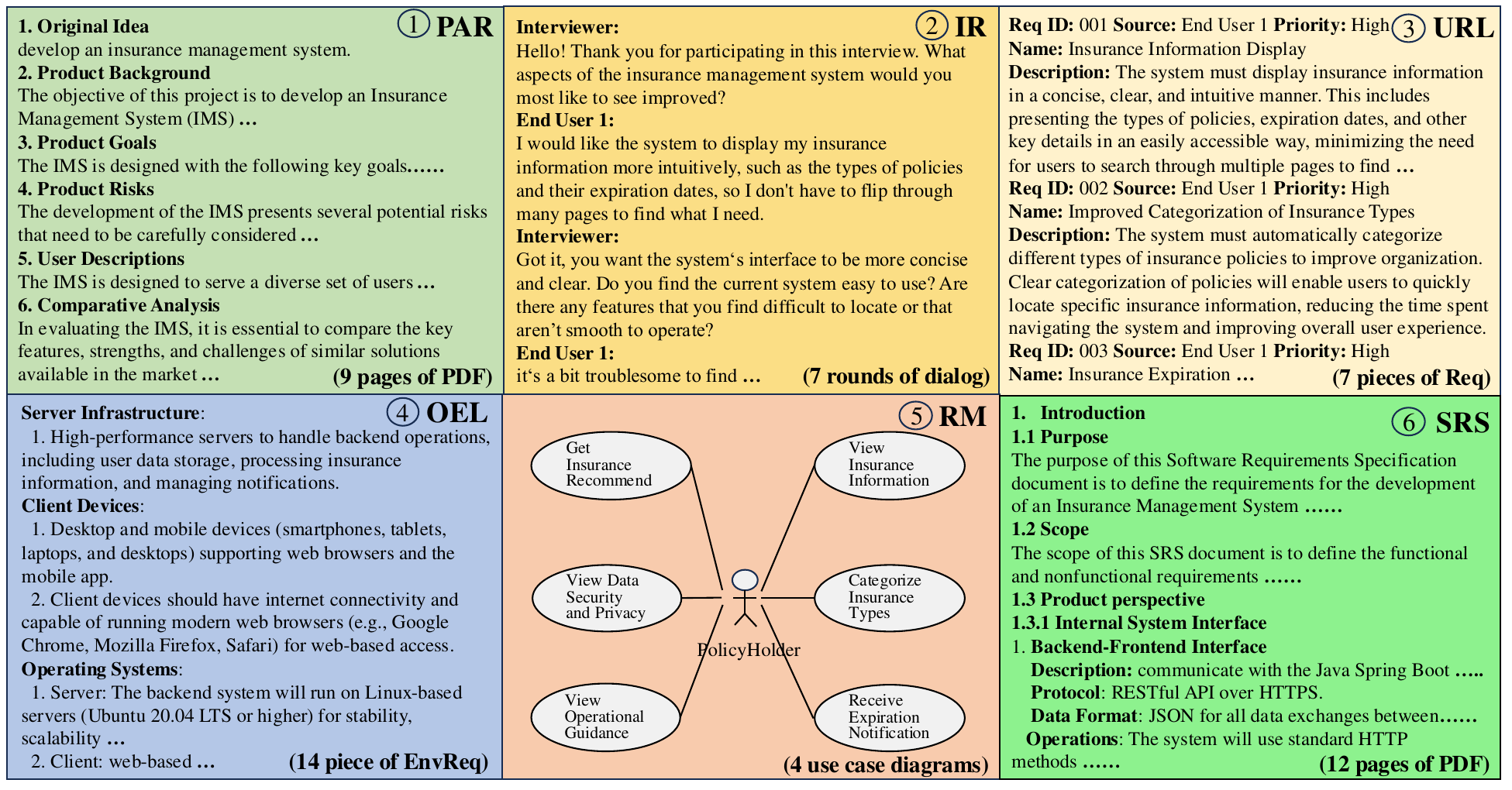}
    \caption{The artifacts generated by our \ours{} on the insurance management system.}
    \label{fig:case study}
\end{figure*}

\subsection{\ours{} Framework}

\textbf{Interviewer.} \ding{182} \textbf{Functionality.} The Interviewer is designed to organize research, collect requirements through interactions with the End-User, and obtain operating environments in collaboration with the Deployer. These processes produce intermediate requirements artifacts such as the Product Analysis Report (PAR), Interview Records (IR), User Requirements List (URL), and Operating Environment List (OEL). \ding{183} \textbf{Knowledge.} To better plan and perform actions, we clarify and integrate various types of knowledge into their prompts, \eg research templates, system-specific knowledge, and requirements templates. 
\ding{184} \textbf{Actions.} \textit{ConductResearch:} The action is designed to search for related questions for the given brief description of the system to be developed, collecting a list of URLs, summarizing the content of the webpages associated with the list, and subsequently writing the PAR. \textit{StartInterview:} The action is to interview End-User to gather requirements. \textit{WriteReqList:} This action involves generating a URL based on the provided IR.

\noindent \noindent \textbf{End-User.} \ding{182} \textbf{Functionality.} The agent collaborates with the Interviewer to provide raw requirement descriptions by responding to the questions posed by the Interviewer. Furthermore, the agent is expected to reflect on and answer more detailed questions, following the guidance provided by the Interviewer, to produce IR. \ding{183} \textbf{Knowledge.} The End-User should possess domain-specific knowledge, scenario-based knowledge, experiential knowledge, and similar expertise to effectively respond to questions from the Interviewer. This knowledge is acquired by simulating potential user requirements and referencing publicly available URL templates from the Internet, among other sources. \ding{184} \textbf{Actions.} \textit{StartRespond:} The action involves responding to the questions provided by the Interviewer and offering a basic description of user requirements, and to generate a raw requirements record corresponding to the given questions. \textit{RespondToRecommend:} This action involves responding to the recommended questions provided by the Interviewer, enabling the End-User to offer additional feedback and clarify their requirements. \textit{RespondToReflection:} This action involves providing detailed explanations and scenario elaborations when the End-User is required to reflect on the current questions posed by Interviewer.

\noindent \textbf{Deployer.} \ding{182} \textbf{Functionality.} The Deployer collaborates with Interviewer to understand the required operating environment conditions for deploying the developed systems and to draft the operating environment documentation. \ding{183} \textbf{Knowledge.} The Deployer gains an understanding of user requirements through knowledge of system configuration, workflows, and system operations. The knowledge is collected from domain-specific sources, such as embedded systems and aerospace fields. \ding{184} \textbf{Actions.} \textit{CheckRunEnvReq: }The action should ensure that the URL aligns with the intended environment to prevent potential runtime issues. \textit{WriteRunEnv:} The action creates the Operating Environment List (OEL) for the system's deployment environment based on the URL.

\noindent \textbf{Analyst.} \ding{182} \textbf{Functionality.} The Analyst first classifies user requirements to select suitable modeling methods. Subsequently, the agent drafts the System Requirements List (SRL), constructs the requirements model (RM), and archives the completed model. \ding{183} \textbf{Knowledge.} The Analyst acquires knowledge of both functional and non-functional requirements, as well as requirements modeling, to classify and prioritize user requirements. The knowledge is provided to the agent through sources such as books, the Internet, and academic papers. 
\ding{184} \textbf{Actions.} \textit{ClassifyUserReqs: } The action differentiates between functional and non-functional requirements and assigns priority levels to these requirements. \textit{WriteSystemReqs: } The action drafts a canonical SRL based on classified and prioritized user requirements to facilitate the construction of RM. \textit{BuildReqModel:} The action identifies requirement entities and their relationships from the SRL and integrates them to construct RM.

\noindent \textbf{Archivist.} \ding{182} \textbf{Functionality.} The Archivist is designed to create Software Requirements Specifications (SRS) based on the provided OEL, SRL, and RM. Additionally, the agent revises the SRS when changes occur in the OEL, SRL, or RM, ultimately producing the final SRS. \ding{183} \textbf{Knowledge.} The Archivist acquires knowledge of SRS templates, which are collected from various domains, standards, and software companies. \ding{184} \textbf{Actions.} \textit{WriteReqSpecs:} The action involves the Archivist drafting the initial version of the SRS based on the OEL, SRL, and RM, while considering the dependency relationships among them. \textit{ ReviseReqSpecs:} The action refines the SRS based on feedback from other agents to ensure its accuracy and alignment with their requirements.

\noindent \textbf{Reviewer.} \ding{182} \textbf{Functionality.} The Reviewer collaborates with Interviewer, Analyst, Deployer, and End-User to review the SRS and provide feedback through written review comments to validate the SRS. \ding{183} \textbf{Knowledge.} Reviewer acquires knowledge of software requirements, domain-specific technical expertise, and structured writing skills to guide the agent in writing comprehensive review comments and generating a valid report. This knowledge is gathered from sources such as the Internet, books, academic papers, and software companies. \ding{184} \textbf{Actions.} \textit{WriteReviewComs:} Review the SRS with the Interviewer, Analyst, Deployer, and End-User, and document the feedback and comments from their review. \textit{WriteValidReport:} The action consolidates reviews from various agents to produce a final set of review comments and feedback, which is then provided to Archivist for revising and refining the SRS. 

%% file: sections/casestudy.tex
\section{Case Study} 
In this section, we present the details of our case study to validate the practicality of our \ours{}, including the selected case, experiment settings, collaboration process, and generated artifacts.

\textbf{The Selected Case.} The selected case is an enterprise-level web-based management project, \ie the insurance management system~\cite{webcase}. 
This system is designed to offer client information management services to insurance companies, primarily for distributing employee welfare insurance and storing related data. We selected this case for two reasons: 1) It presents a widely applied business scenario in the real world and involves various requirements-related tasks; 2) It includes multiple different functional modules with relatively complex requirements, demonstrating the application of our \ours{} in real-world scenarios.

\textbf{Experiment Settings.} We use \textit{GPT-4-turbo-2024-04-09}~\cite{openai2023gpt} as the base LLM of our \ours{}. It is an autoregressive language model developed by OpenAI \cite{web:openai}. To ensure the stability of the generated artifacts, the output confidence parameter \textit{Top-P} was set to 1.0 using the default value, and the frequency and presence penalties were set to 0.0. The output randomness parameter \textit{Temperature} was set to 0.3, and the \textit{maximum token size} was set to 4096. Besides, we set the artifacts pool empty. 

\textbf{Collaboration Process.} First, we manually input a brief description of the insurance management system, \ie \textit{I want to develop an insurance management system.} This description is added to the artifacts pool. Subsequently, the Interviewer observes the addition of this artifact to the pool and selects the \textit{WriteProductAnalysisReport} action to perform from their predefined action list. This action generates a product analysis report and put it to the pools. Next, the Interviewer notices that a product analysis report has been added to the pool and chooses to execute the \textit{PrepareInterviewList} action, generating an interview question list, which is added to the pool. At this point, the Interviewer and the End-User engage in 7 rounds of dialogue to gather user requirements. After the dialogue concludes, the Interviewer executes the \textit{WriteReqList} and \textit{WriteInterviewRecord} actions, adding the User Requirements List and Interview Record to the artifact pool. Next, the Deployer observes the addition of the user requirements list to the pool and executes the \textit{WriteRunEnv} action to add the Operating Environment List artifact. Then, the Analyst notices the presence of both the User Requirements List and the Operating Environment List in the pool and proceeds to execute the \textit{WriteSystemReqs} action, adding the System Requirements List to the pool. Upon noticing the addition of the System Requirements List, the Analyst executes the \textit{BuildReqModel} action, generating four use case diagrams, which are then added to the artifact pool. Finally, the Archivist begins the process of writing the requirements specification, adding the Software Requirements Specification to the pool. The Reviewer and other agents then engage in four rounds of dialogue and write various review comments based on the conversation records. 

\textbf{Generated Artifacts.} Figure~\ref{fig:case study} presents partial content of the six major artifacts generated in this case study. It can be observed that the quality of both the intermediate artifacts and the final SRS, generated through the collaboration of multiple agents within our \ours{} framework, is satisfactory.
The complete set of artifact contents can be accessed via our public link~\cite{web:code}.

%% file: sections/future_work.tex
\section{Future Work Roadmap}
\label{sec:future_work}
To advance our vision of improving requirements development, we highlight several significant research opportunities and discuss the benefits of automating this process.

\textbf{Integration of Knowledge-Driven Reasoning.} Integrating advanced knowledge-driven reasoning presents significant research opportunities for enabling agents to make informed decisions in complex scenarios. This approach relies on domain expertise and emphasizes the simulation of expert reasoning processes, enabling agents to perform more efficient reasoning that facilitates planning, decision-making, and action. Exploring the opportunity can push the boundaries of current requirements engineering practices and lay the groundwork for broader applications of knowledge-driven reasoning in multi-agent systems.

\textbf{Autonomous and Collaborative Multi-Agent Framework.}  Autonomous agents can automate requirement elicitation, classification, analysis, modeling, and validation, significantly reducing human intervention while enhancing overall efficiency. Collaborative mechanisms enable agents to work in a coordinated manner, minimizing redundancy and ensuring comprehensive task execution. Requirements development involves multiple stakeholders and complex interaction logic, collaborative agents address these complexities by decomposing them into manageable tasks through effective collaboration in large-scale projects. Additionally, information sharing and dynamic cooperation among agents facilitate faster adaptation to the artifact changes, improving the management of complex or cross-domain requirements development.

\textbf{Automated Feedback Loop for Enhancing Traceability.}
Automated feedback loop can significantly enhance traceability by continuously monitoring the relationships between requirements, design artifacts, and implementation components. When a new task or artifact is introduced into the artifacts pool, it triggers the feedback loop, prompting the corresponding agent to take action and complete its assigned function. Each artifact adheres to its dependency relationships within the system, and the mechanism reduces the risk of errors and facilitates compliance with regulatory standards and project goals. Furthermore, it supports iterative development by ensuring that changes in requirements are seamlessly propagated across the system, maintaining alignment between stakeholder needs and system functionality.